\documentclass[]{aa}
\usepackage{graphicx}
\begin{document}
\title{Investigating star formation in the young open cluster
NGC~6383\thanks{Based on 
observations at ESO-La Silla (Proposal 073.C-0144)}}
\author{E.~Paunzen, M.~Netopil, K.~Zwintz}

\mail{Ernst.Paunzen@univie.ac.at}

\institute{Institut f\"ur Astronomie der Universit\"at Wien,
           T\"urkenschanzstr. 17, A-1180 Wien, Austria}

\date{Received 2006; accepted 2006}

\abstract{ By studying young open clusters, the mechanisms important for 
star formation over several Myr can be examined. For example, accretion  
rate as a function of rotational velocity can be investigated.  Similarly, sequential star formation 
triggered by massive stars with high mass-loss rates can be studied in detail. }
{ We identified and characterized probable members of NGC~6383, as well as determined cluster parameters. }
{ New Str{\"o}mgren $uvby$ CCD photometry, obtained by us, is presented.  This new data, 
together with Johnson $UBV$ and 2MASS data in the 
NIR, was used to investigate characteristics of pre- as well as zero age main sequence cluster members. }
{ We present Str{\"o}mgren $uvby$ CCD photometry for 272 stars in the field of NGC~6383
and derive its reddening, $E(b-y)\,=\,0.21(4)$\,mag, as well as distance, 
$d$\,=\,1.7(3)\,kpc from the Sun. Several stars with NIR excess and objects in the
domain of the classical Herbig Ae/Be and T Tauri stars were detected. Two previously known
variables were identified as rapidly-rotating PMS stars. The field population is
clearly separated from the probable members in the color-magnitude diagram. }
{ NGC~6383 is a young open cluster, with an age of less than 4 Myr, undergoing 
continuous star formation. True pre-main sequence members might be found down
to absolute magnitudes of +6\,mag, with a variety of rotational velocities
and stellar activities. }


\maketitle

\section{Introduction}

Young open clusters offer the opportunity to investigate star formation simultaneously 
for a significant number of stars from high to very low mass. Processes like accretion,
mass-loss and pulsation in the presence of both local and global magnetic fields can
be studied (Monin et al. 2006). The pre-main sequence (PMS) phase of 
stellar evolution has been extensively investigated in the
last decade (James et al. 2006). With the detection of magnetic fields (Wade et al. 2005) and the
application of asteroseismic tools (e.g. Ripepi et al. 2006, ; 
Zwintz, Guenther \& Weiss 2006) the early phases of the stellar evolution 
begin to reveal their mysteries. 

We investigate the southern young open cluster NGC~6383 located at
$\alpha$(2000.0)\,=\,17$^{\rm h}$34$^{\rm m}$48$^{\rm s}$,
$\delta$(2000.0)\,=\,$-$32$\degr$34$\arcmin$00$\arcsec$; $l$\,=\,355$\degr$690, $b$\,=\,0$\degr$041 
which has been a target of
several investigations (see Zwintz et al. 2005 for a summary). This aggregate is especially interesting
because it contains several variable stars and, due to a very massive binary system in its core,
star formation is still ongoing (Rauw et al. 2003). 

We present new Str{\"o}mgren $uvby$ photometry of several fields of NGC~6383,
together with a detailed analysis of the available 2MASS data (Skrutskie et al. 2006).
The color-magnitude diagram clearly shows continuous star formation over about the last 4\,Myr. We
have derived the cluster parameters and discovered two rapidly-rotating PMS stars.  These are
very interesting objects for further observations. An analysis of the NIR data shows that
PMS members are present to spectral types of M0 or even cooler. Some members show
a NIR excess and are located in the photometric domain of Herbig Ae/Be and classical T Tauri stars.
Based on all available information from our observations and the literature, 
a list of bona-fide members of NGC~6383 was compiled.   

\begin{figure}
\begin{center}
\includegraphics[width=85mm]{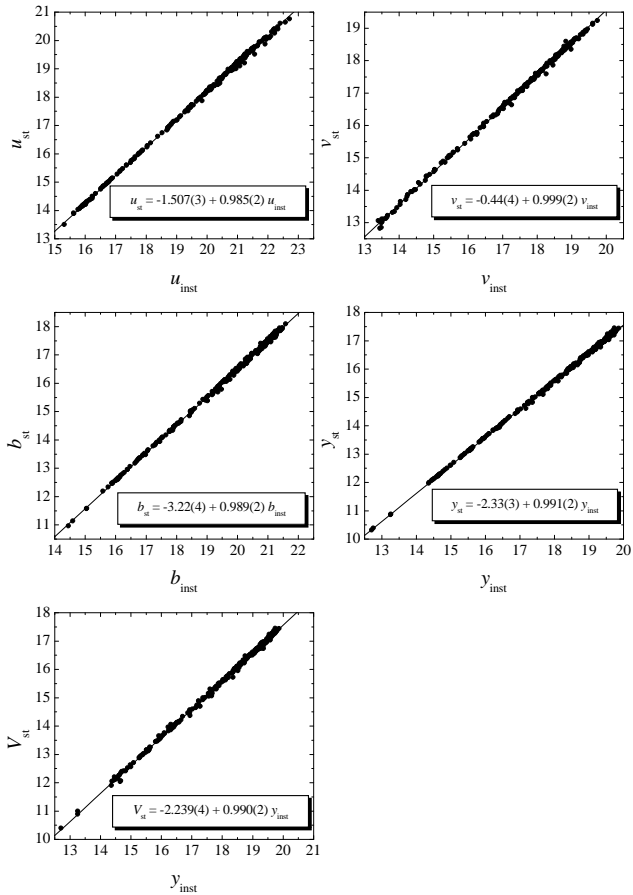}
\caption[]{The standard relations of our instrumental magnitudes (suffixes ``inst'') 
to the highly accurate $uvby$ photometry of IC 4651 published by Meibom (2000).}
\label{ic4651}
\end{center}
\end{figure}

\begin{figure}
\begin{center}
\includegraphics[width=95mm]{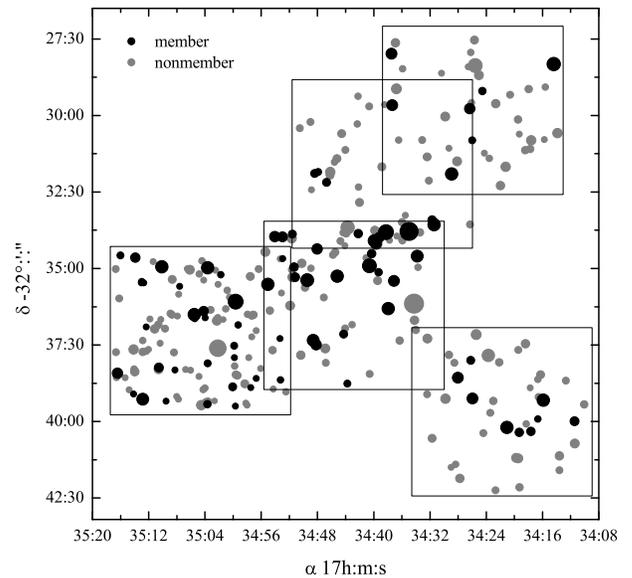}
\caption[]{The five observed fields in NGC~6383.
The sizes (by area) of the symbols are inversely proportional to the $V$-magnitudes.}
\label{chart}
\end{center}
\end{figure}

\begin{figure}
\begin{center}
\includegraphics[width=80mm]{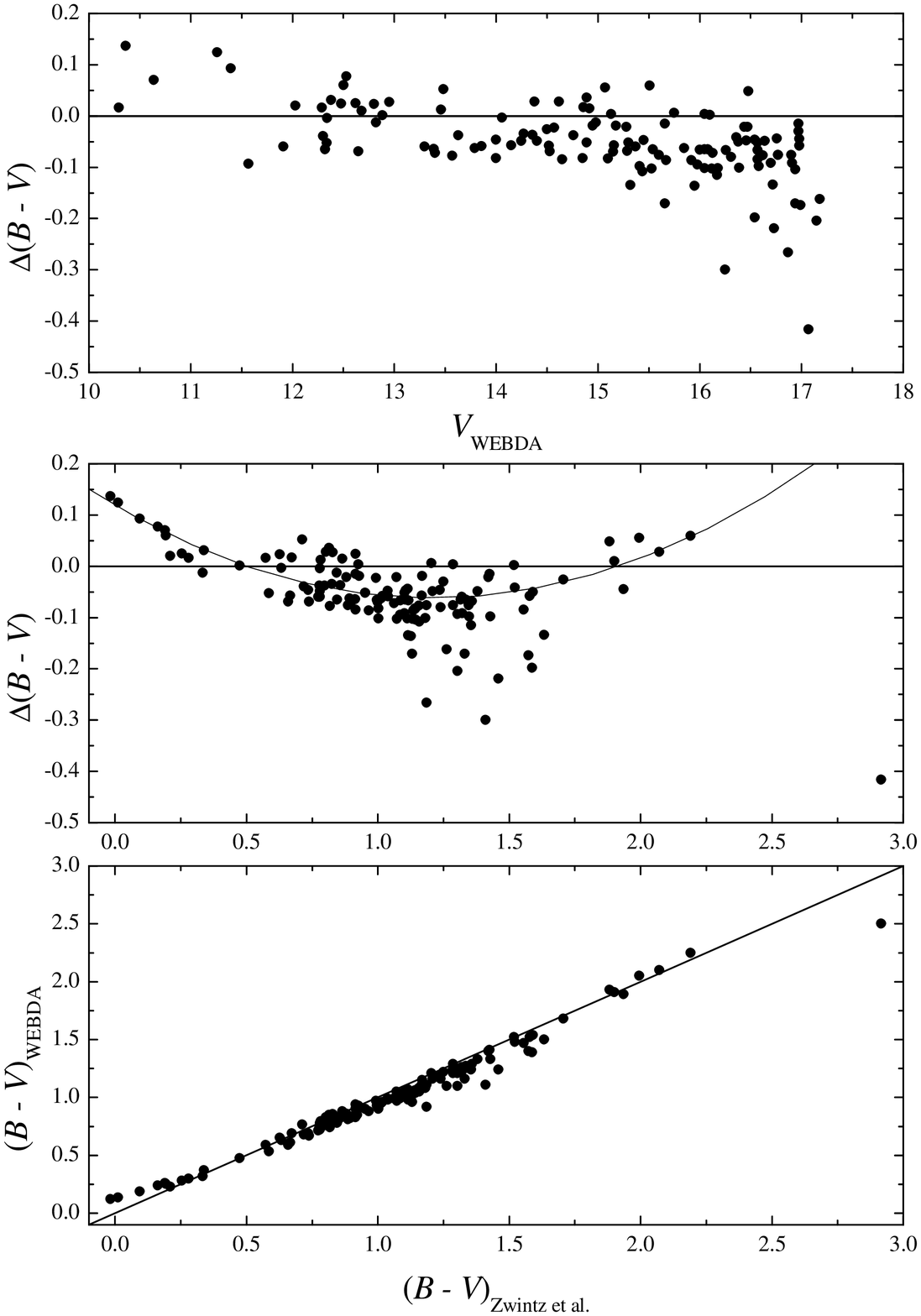}
\caption[]{Johnson $BV$ photometry by Zwintz et al. (2005) in comparison
with the mean photometric values from WEBDA. The mean $\Delta (B-V)$\,=\,$(B-V)_W$\,$-$\,$(B-V)_Z$
for the intermediate region
(0.5\,$<$\,(B-V)\,$<$\,1.5) is about 0.07\,mag, whereas for hotter and cooler stars there is a
systematic trend which can be described by a second-order polynomial.}
\label{stanzl}
\end{center}
\end{figure}

\section{Observations and reduction}

The observations were performed during the night of 13./14.06.2004 at the 
3.6\,m telescope (ESO-La Silla), with the EFOSC2 (fast modus) and the Loral/Lesser 2048\,$\times$\,2048 pixel
(1\,pixel\,=\,0$\farcs$157) CCD, which yields a 5$\arcmin$ field-of-view using a
standard Str{\"o}mgren $uvby$ filter set (Observer: M.~Netopil). 

As NGC~6383 is believed to be rather extended, about 20$\arcmin$, we have chosen five overlapping
fields in the cluster area to cover the most interesting objects. In total, 33 frames in
each filter were observed with integration times ranging from 5 to 300 seconds.

The bias-subtraction, dark-correction, flat-fielding and point-spread-function fitting 
were carried out within standard IRAF V2.12.2 routines. 

No offsets in instrumental magnitudes between the different overlapping
fields were detected, within the photometric and transformation error limits. 

The accurate $uvby$ photometry of IC~4651 published by Meibom (2000) was
used to transform our instrumental magnitudes to standard ones. We have observed
IC~4651 at different airmasses resulting in 25 individual frames. After correcting
all frames for the airmass, the following standard relations were
derived (Fig. \ref{ic4651}). The errors in the final digits of the corresponding quantity
are given in parentheses.
\begin{eqnarray}
u_{st} &=& -1.507(3) + 0.985(2) u_{inst} \\
v_{st} &=& -0.441(3) + 1.000(2) v_{inst} \\
b_{st} &=& -3.220(3) + 0.986(2) b_{inst} \\
y_{st} &=& -2.262(3) + 0.991(2) y_{inst} \\
V_{st} &=& -2.239(4) + 0.990(2) y_{inst}
\end{eqnarray} 
Those relations were applied to the instrumental magnitudes of the stars in the
five fields of NGC~6383. 

For a final test, we compared our $uvby$ photometry with data from Eggen (1978) for six stars
in both our sample and his. He used a narrower $v$ filter which
only effects $m_1$ and $c_1$. The values agree very well within the error limits. 

In total, we observed 272 stars, of which 105 could not be measured in
Str{\"o}mgren $u$ because of their relative faintness and the limited  
quantum efficiency of the instrument.

The table with all the data for individual cluster stars, as well as
nonmembers, is available
in electronic form at the CDS via anonymous ftp to cdsarc.u-strasbg.fr (130.79.128.5),
http://cdsweb.u-strasbg.fr/Abstract.html
or upon request from the first author. The table includes our internal
numbers, the WEBDA and 2MASS identifiers,
J2000.0 coordinates, the $X$ and $Y$
coordinates within our frames, the observed $(b-y)$, $m_1$ and
$c_1$ values with their corresponding errors, $V$ magnitudes,
the $(B-V)$ colors from the literature and the number of observations, respectively.

\begin{figure}
\begin{center}
\includegraphics[width=80mm]{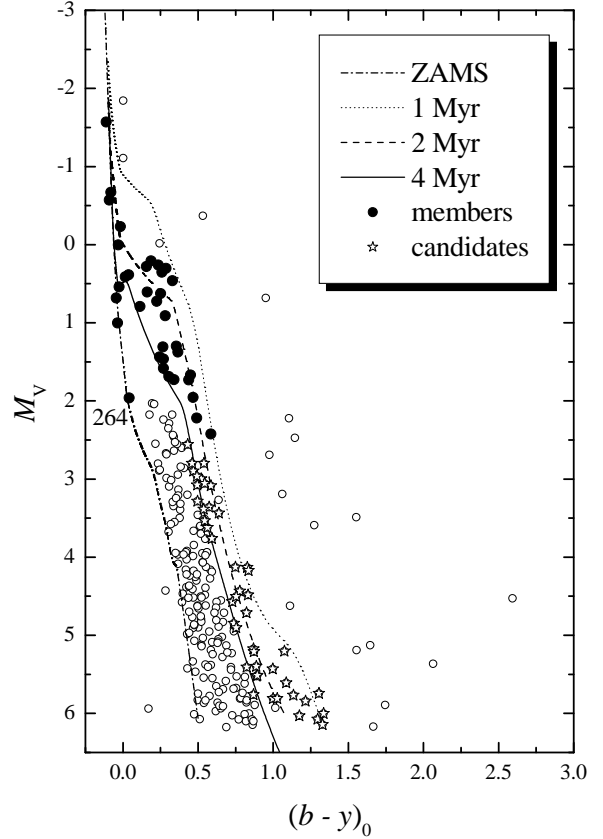}
\caption[]{Color-magnitude diagram assuming $E(b-y)\,=\,0.21$\,mag and 
$V - M_V$\,=\,12.0\,mag. The zero age main sequence is taken from Philip \& Egret (1980)
whereas the pre-main sequence tracks are from Siess et al. (2000). Non-members brighter than
+1\,mag were selected from the literature (e.g. FitzGerald et al. 1978). }
\label{mvby}
\end{center}
\end{figure}

\begin{figure}
\begin{center}
\includegraphics[width=80mm]{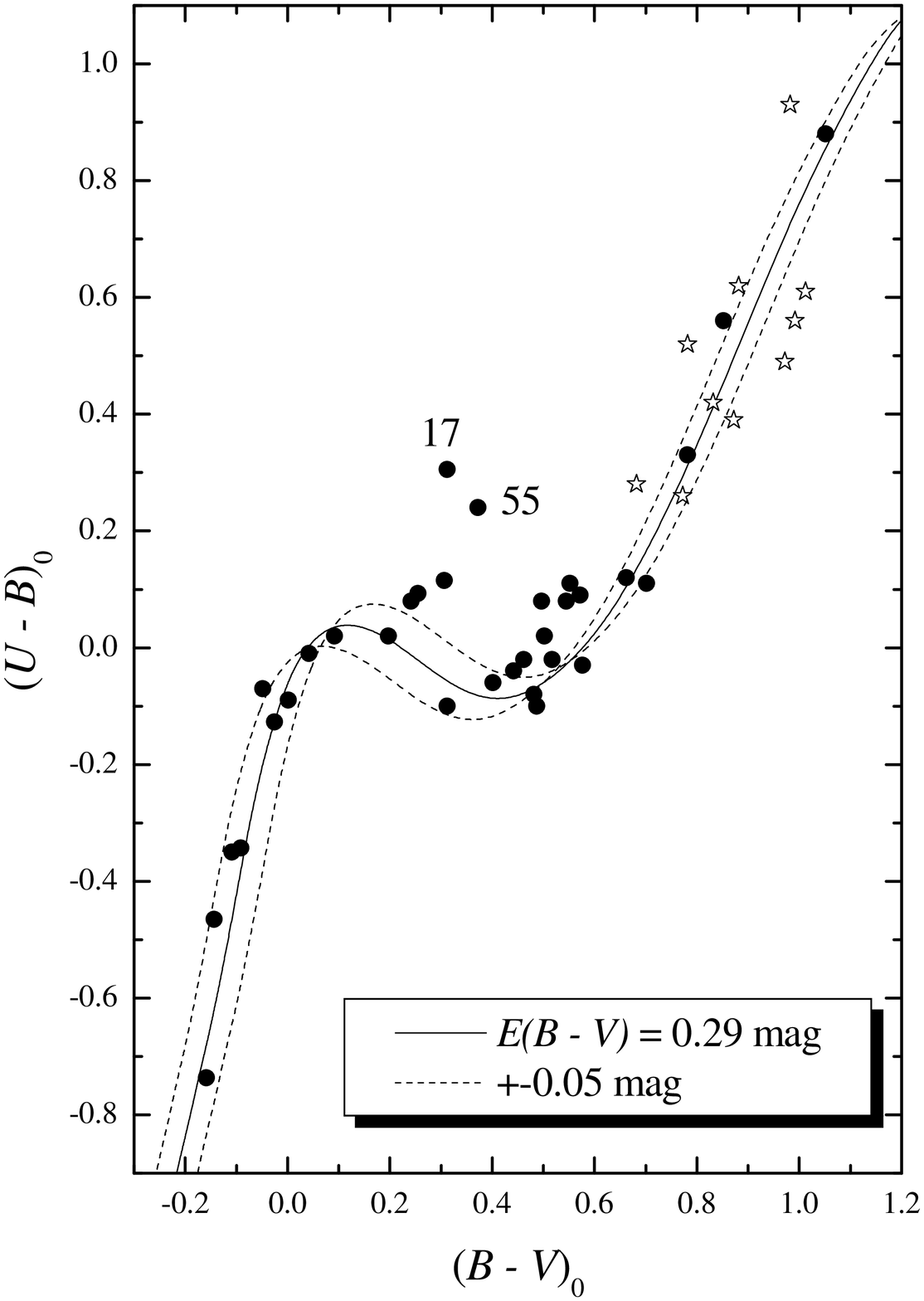}
\caption[]{$(U-B)_0$ versus $(B-V)_0$ diagram for the members (filled circles)
and possible members (asterisks). The standard line is taken from Girardi et al. (2002)
computed for $E(B-V)\,=\,0.29$\,mag and an error limit
of $\pm$0.05\,mag. The only significant outliers are the stars No.
17 and 55 (known as variable).}
\label{ubbv}
\end{center}
\end{figure}

\section{Results}

Throughout this paper we have adopted the numbering system from WEBDA (http://www.univie.ac.at/webda), 
only in some cases we also list the numbers according to FitzGerald et al. (1978, F\#) 
and Zwintz et al. (2005, Z\#).

NGC~6383, which is associated with a large H\,{\sc II} region,
was targeted for several investigations in the past. They have been summarized
by Zwintz et al. (2005) in some detail. The basic parameters of this cluster,
from different sources, are: $E(B-V)$\,=\,0.31(3)\,mag, 1.4\,$<$\,$d$\,$<$\,2.5\,kpc and
log\,$t$\,$<$\,4\,Myr. Its apparent diameter is still a matter of
debate. FitzGerald et al. (1978) concluded on the basis of star counts (see Figure 1 therein)
that only a core radius of 2$\arcmin$ should
be considered, whereas Zwintz et al. (2005) give a value of 10$\arcmin$, which is based on
Lyng\aa\,(1987). Kharchenko et al. (2005) list a core and cluster radius of 4$\farcm$8,
and 15$\arcmin$, respectively. Their age estimate of 5\,Myr is compatible with the literature
whereas their distance of 985\,pc seems too low.

Figure \ref{chart} shows the observed five fields with the location of members
and non-members. The sizes (by area) of the symbols are inversely proportional to the 
$V$-magnitudes. Inspecting the area surrounding NGC~6383 in the Digitized Sky Survey
(DSS), one can see several empty areas, probably caused by dark clouds. 
Dobashi et al. (2005) have also listed that region in their Atlas and 
Catalog of Dark Clouds, reporting a cloud and clumps extending over 0.27$\degr$ in this area.  
On the basis of this fact, a density profile using, for example, the extensive 
2MASS data will evoke erratic results for the cluster diameter. However, 
a cluster radius of 10$\arcmin$ to 15$\arcmin$ seems to be within 
the realm of possibility, since the confirmed members are distributed 
over the whole observed field of $\sim$\,15$\arcmin$ (Fig. \ref{chart}).
From the comparison of Fig. \ref{chart} with the DSS images, we conclude that
we have covered only 60\% of the complete cluster area, introducing an
additional bias. Only further photometric observations can help 
establishing the diameter of NGC~6383.

All members of NGC~6383 which are later than a spectral type of A0 are
still in their PMS phase. Rauw et al. (2003) studied the cluster in X-rays 
using data from the XMM-Newton satellite. They detected several X-ray sources, which were 
interpreted as signs of young T Tauri stars. 
FitzGerald et al. (1978) concluded that the central
bright ($V$\,=\,5.7\,mag) spectroscopic binary (O7\,V + O7\,V) system, HD~159176 (WEBDA No. 1),
triggered star formation within its surroundings. They derive an age of 2.8(5)\,Myr
for it.

No estimation of the cluster metallicity exists in the literature.
There are neither spectroscopic abundance investigations nor a photometric 
determination available. The latter is caused by the fact that only stars
hotter than A0 are on the zero age main sequence and no giant or supergiant
is a member of NGC~6383. Photometric metallicity calibrations are only available
for giants (Hilker 2000) or stars later than A0 (Karaali et al. 2005). A query
in WEBDA reveals that there are no other open clusters in the galactic vicinity of NGC~6383
with a known metallicity. The galactic distance, $R_{GC}$, of NGC~6383 is about 5.0 to 6.0\,kpc
depending on its true distance from the Sun ($R_0$\,=\,8.0\,kpc). Taking a mean
abundance gradient of $-$0.04 dex\,kpc$^{-1}$ (Cunha \& Daflon 2005) for the Milky
Way, an upper limit for the metallicity of $-$0.12\,dex is evident. Such a small
difference from the solar value does not alter the isochrone significantly 
(Girardi et al. 2002). However, it has to be emphasized that the ``intrinsic''
range of metallicities for a constant $R_{GC}$ is about $\pm$0.25\,dex (Chen et al. 2003), but
data for open clusters in the direction to the galactic center is still
very rare. 

\begin{table*}
\caption{ A list of the 44 probable members of NGC~6383 that are still in their
PMS phase. These objects were selected on the basis of several photometric
diagrams. The column ``Flag'' denotes stars that are in the PMS region
in the NIR diagram (Fig. \ref{hkjh}). }
\begin{center}
\begin{tabular}{rcccc|rcccc}
\hline
\hline
\multicolumn{1}{c}{$No(our)$} & \multicolumn{1}{c}{$No(WEBDA)$} &  
\multicolumn{1}{c}{$(b-y)_0$} & \multicolumn{1}{c}{$M_V$} & \multicolumn{1}{c|}{Flag} &
\multicolumn{1}{c}{$No(our)$} & \multicolumn{1}{c}{$No(WEBDA)$} &  
\multicolumn{1}{c}{$(b-y)_0$} & \multicolumn{1}{c}{$M_V$} & \multicolumn{1}{c}{Flag} \\
\hline
8	&	348	&	0.823	&	4.71	&	*	&	163	&	538	&	0.997	&	5.43	\\
13	&	222	&	1.134	&	5.77	&		&	164	&	335	&	1.027	&	5.81	\\
21	&	 	&	0.868	&	5.18	&		&	169	&	 	&	0.868	&	5.76	\\
23	&	 	&	0.872	&	5.21	&	*	&	178	&	596	&	0.563	&	3.62	\\
26	&	 	&	0.890	&	5.52	&	*	&	179	&	607	&	0.520	&	3.00	\\
28	&	206	&	0.999	&	5.81	&		&	180	&	611	&	0.539	&	2.80	\\
36	&	444	&	0.461	&	2.80	&		&	191	&	378	&	0.545	&	3.54	\\
44	&	 	&	1.087	&	5.61	&	*	&	193	&	379	&	0.738	&	4.85	&	*	\\
55	&	 	&	1.216	&	5.84	&		&	197	&	382	&	0.574	&	3.36	&	*	\\
61	&	343	&	0.894	&	5.52	&	*	&	206	&	251	&	0.537	&	3.39	\\
73	&	199	&	1.074	&	5.20	&	*	&	211	&	535	&	0.728	&	4.57	\\
86	&	 	&	1.296	&	6.07	&		&	217	&	388	&	0.494	&	3.08	&	*	\\
87	&	 	&	0.828	&	4.13	&		&	226	&	604	&	0.585	&	3.09	\\
90	&	543	&	0.778	&	4.44	&		&	231	&	609	&	0.748	&	4.13	\\
106	&	337	&	0.824	&	5.41	&		&	273	&	128	&	0.592	&	3.76	\\
116	&	541	&	0.639	&	3.44	&	*	&	277	&	503	&	0.544	&	3.06	\\
118	&	546	&	0.832	&	4.49	&	*	&	297	&	313	&	0.494	&	2.97	\\
119	&	 	&	1.335	&	6.00	&	*	&	301	&	571	&	0.753	&	4.50	\\
120	&	 	&	1.175	&	6.03	&	*	&	307	&	566	&	0.837	&	4.17	\\
126	&	134	&	0.888	&	5.40	&	*	&	308	&	312	&	0.497	&	3.29	\\
136	&	539	&	1.305	&	5.75	&		&	312	&	310	&	0.753	&	4.91	\\
143	&		&	1.328	&	6.14	&		&	328	&	309	&	0.432	&	2.56	\\
\hline
\hline
\label{probable}
\end{tabular}
\end{center}
\end{table*}

\subsection{Optical properties}

For the further analysis,
we have checked the intrinsic consistency of the available Johnson $UBV$ photometry included
in WEBDA. In total, twelve independent data sources are available. The only
systematic offset was found for the $(B-V)$ data 
by Zwintz et al. (2005). One has to 
keep in mind that the cited photometry was not obtained for an 
astrophysical analysis of cluster properties, but to detect pulsating 
members. Hence, the reduction was tuned for high relative photometric 
accuracy on time scales of hours. Not surprisingly, a re-reduction of 
the data in conformity with Johnson standards arrives at slightly 
different values. Using the method described in Mermilliod \& Paunzen 
(2003), a second-order polynomial fit to the deviations (middle panel of 
Fig. \ref{stanzl}) provides an intrinsically more accurate transformation to the 
Johnson system. We have not included these data in our further 
analysis.

The PMS evolutionary tracks by Siess et al. (2000) were fit 
to the cluster's color-magnitude diagram, and used to determine the 
cluster's distance and reddening.  
These tracks are listed for luminosity and effective 
temperature, respectively. The transformation of luminosity into absolute
magnitude is straightforward. More of a problem is the transformation of 
effective temperature into Str{\"o}mgren $(b-y)$ because several calibrations
are available. First of all, the empirical ZAMS listed by 
Philip \& Egret (1980) was used to convert the ends of the PMS tracks into the $(b-y)$ and $M_V$ 
plane. We found that the calibration by Hauck \& K{\"u}nzli (1996, Table 1)
satisfies the color-magnitude diagram best. The results by Clem et al. (2004) confirm
our choice. The final transformation is given as
\begin{eqnarray}
\Theta_{eff} &=& +0.564(8) + 2.961(23) (b-y)_0 \quad (b-y)_0 < 0 \\
\Theta_{eff} &=& +0.564(9) + 0.822(3) (b-y)_0 \quad (b-y)_0 > 0 \\
\Theta_{eff} &=& \frac{5040\,K}{T_{eff}}
\end{eqnarray} 
including a small shift of +0.02\,mag compared to the original equations
to match the standard ZAMS line by Philip \& Egret (1980).

Figure \ref{mvby} shows the color-magnitude diagram with an 
adopted reddening of $E(b-y)\,=\,0.21$\,mag and a distance modulus 
$V - M_V$ of 12.0\,mag. This corresponds to $E(B-V)\,=\,0.29$\,mag and
a distance of about 1.7\,kpc. The objects marked with filled circles are
members according to the literature, asterisks 
represent the possible PMS members taken from Table \ref{probable}. 
The stars with absolute magnitudes
brighter than +1\,mag are identified non-members taken from the literature 
(e.g. FitzGerald et al. 1978). 
Figure \ref{ubbv} shows
the $(U-B)_0$ versus $(B-V)_0$ for cluster members, with the standard line from 
Girardi et al. (2002), computed for $E(B-V)\,=\,0.29$\,mag and an error limit
of $\pm$0.05\,mag. This error band represents the main sequence up to
$(B-V)_0$\,=\,+0.1\,mag very well. Cooler objects are still in their PMS 
phase and can have a peculiar behavior in this diagram. The $\pm$0.05\,mag 
error corresponds
to $\Delta E(b-y)$\,=\,0.036 and $\Delta A_V$\,=\,0.155\,mag. In Fig. \ref{ubbv},
two outliers are noticeable, the objects No. 17 and 55. These two stars show
no other abnormal behavior, which prevents any reliable conclusion about their
nature. The star No. 55 is variable according to Zwintz et al. (2005) with an amplitude
of about 25\,mmag, which is too small to explain the apparent shift. 

\begin{figure}
\begin{center}
\includegraphics[width=80mm]{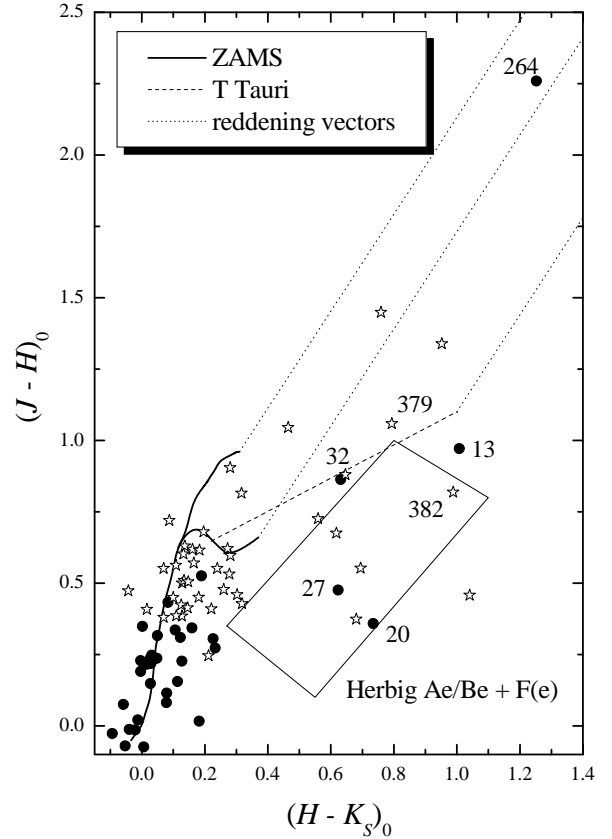}
\caption[]{$(J-H)_0$ versus $(H-K_S)_0$ diagram.
The standard lines for the main sequence are from Bessell \& Brett (1988), with the reddening
vectors (dotted lines) according to Rieke \& Lebofsky (1985). The locus of the classical T Tauri stars 
(dashed line) is from Meyer et al. (1997), whereas the region of Herbig Ae/Be and emission line
F-type objects is taken from Hernandez et al. (2005).}
\label{hkjh}
\end{center}
\end{figure}

The observed color-magnitude diagram (Fig. \ref{mvby}) shows very interesting
features. First of all, there is a lack of objects with absolute magnitudes between
+0.5 and +2.0\,mag. The only exception is object No. 264, for which a large reddening
in the NIR is evident (Fig. \ref{hkjh}). This gap seems to be the borderline between the
members of NGC~6383 and the field population. The members in the PMS phase lie all within
the isochrones between 1 to 4\,Myr down to $M_V$\,=\,2.5\,mag (F0). These are
PMS A-type objects mixed together with Herbig Ae/Be stars (Hernandez et al. 2005). The classical
T Tauri stars have spectral types later than G0 (Meyer et al. 1997) which corresponds to 
$M_V$\,$\approx$\,4.3\,mag.

The estimation of 
distance error is difficult because neither a turn-off point nor red
giants are present in NGC~6383. That is the reason why widely different distance
estimations (1.4 to 2.5\,kpc which corresponds to $\Delta M_V$\,$\approx$\,1.1\,mag
for a constant reddening) for this aggregate can be found in the literature.
From Fig. \ref{mvby} we are able to make a rough estimation of the distance error
by inspecting the location of members with respect to the ZAMS and PMS tracks.
Taking into account the error in reddening for both parameters we can still fit all features,
within observational errors, with a shift of $\pm$0.4\,mag.  The final cluster parameters and 
their errors are: $E(B-V)\,=\,0.29(5)$\,mag and $d$\,=\,1.7(3)\,kpc.

\subsection{Near-Infrared properties}

The 2MASS data base (Skrutskie et al. 2006) with Near-Infrared measurements (NIR) was queried for
appropriate entries in a radius of 10$\arcmin$ around the cluster's center, covering the five
observed fields. The coordinates of these objects were then transformed into a rectangular 
(X,Y) frame and compared with the star positions within our images. We have only included objects
with 2MASS ($JHK$) measurements within two arc seconds of the optical source and with unambiguous 
identifications.  This resulted in 262 positive detections.

We have used the 2MASS data to search for signs of NIR excess and circumstellar dust. 
The results of the method described by Suchkov et al. (2002), 
analyzing the $(V-K_S)$ versus $(b-y)_0$ diagram, are in total agreement with the much more accurate
conclusions from the $(J-H)_0$ versus $(H-K_S)_0$ diagram, shown in Fig. \ref{hkjh}, for
all bona-fide and possible members. 

The standard lines for the main sequence included in Fig. \ref{hkjh} are from Bessell \& Brett (1988) 
with the reddening
vectors according to Rieke \& Lebofsky (1985). The differences between the 2MASS photometric system
and the one from Bessell \& Brett (1988) are only marginal, in the range of a few
hundredths of a magnitude (Carpenter 2001, Appendix A). The locus of the classical T Tauri stars 
is from Meyer et al. (1997), whereas the region of Herbig Ae/Be and emission line
F-type objects is taken from Hernandez et al. (2005). Notice that the weak-emission T Tauri stars 
can not be separated from main sequence objects (Meyer et al. 1997). We find
five objects with a significant NIR-excess:
\begin{itemize}
\item Stars with known NIR excess: Nos. 20, 27 (F4) and 264 (F6) were targets of the investigation
by van den Ancker et al. (2000). The amount of excess nicely correlates with the deviation
from the standard relation for normal type stars within our diagram (see their Fig. 2).
\item New detections: No. 13 and 32, both show a moderate excess, which makes them interesting
targets for a spectroscopic investigation.
\end{itemize}

The star No. 264 is highly reddened, probably caused by circumstellar material. There are also
a few probable members which show similar but less extreme behavior.

For the final step, we have investigated all cool stars located inside the PMS tracks (Fig. \ref{mvby})
for which 2MASS measurements are available.
Fourteen of them are placed in the domain of Herbig Ae/Be and T Tauri stars, with thirty additional 
objects located in the region of the standard relation (Fig. \ref{hkjh}). 
Ten objects have available $(U-B)$ measurements (Fig. \ref{ubbv}), placing them well on the standard relation
given the cluster's reddening. Table \ref{probable} lists the 44 objects which are good candidate 
PMS members of NGC~6383, selected on the basis of the above mentioned criteria. 
Further spectroscopic and kinematic data are required to establish their nature and membership.

\subsection{Two rapidly rotating PMS stars}

We would like to comment on two very interesting stars, No. 379 (Z64) and 382 (Z71), 
published by Zwintz et al. (2005), for which our new photometry adds important information. 
Both stars fall well within the PMS domain in the NIR (Fig. \ref{hkjh}) and can be regarded
as members of NGC~6383. This is supported by the reddening-free [$m_1$] values of 0.414 and 
0.525\,mag respectively, which are typical for stars later than K (Str{\"o}mgren 1966). The 
hypothesis that both are highly reddened B-type stars, as formulated by Zwintz et al. (2005),
can therefore be excluded. The angular velocities of both objects are about 16 radians\,d$^{-1}$
or approximately 30 to 40\% of their break-up velocity, 
if we assume that the found periods are due to rotation. Such high values are not 
typical for PMS stars (Herbst et al. 2002) but have been observed in similar 
objects (Strassmeier et al. 2005). Rapidly-rotating PMS stars are most important
for understanding the accretion process combined with the effects of stellar magnetic
fields (Eisner et al. 2005).

\section{Conclusions}

We have investigated the young open cluster NGC~6383 with several photometric
systems, presenting new Str{\"o}mgren $uvby$ CCD photometry for 272 stars.  
From this data we derive a reddening of $E(b-y)\,=\,0.21(4)$\,mag
and a distance of $d$\,=\,1.7(3)\,kpc. An upper age limit of 4\,Myr was determined for NGC~6383. 
Neither a turn-off point nor red giants have been detected.

In the $M_V$ versus $(b-y)_0$ color-magnitude diagram, 44 probable PMS members could
be traced from A to very cool M-type objects. At least 14 of them are in the NIR (2MASS)
domain of classical T Tauri stars. Appropriate isochrones clearly show that members
are present from 1 to 4\,Myr with a distinct separation from the field population at
absolute magnitudes between +0.5 and +2.0\,mag. 

Five stars with a large NIR excess are unambiguously identified in a  
$(J-H)_0$ versus $(H-K_S)_0$ diagram.

We also report the identification of two rapidly-rotating PMS stars with
angular velocities of approximately 30 to 40\% of their break-up velocity.
Such objects are still very rare and most important for out understanding of the early
stages of stellar evolution, in the presence of accretion and local magnetic
fields.

\begin{acknowledgements}
We would like to dedicate this paper to Hartmut Holweger who died during its
preparation. This research was performed within the projects  
{\sl P17580} and {\sl P17920} of the Austrian Fonds zur F{\"o}rderung der 
wissen\-schaft\-lichen Forschung (FwF). K.~Zwintz acknowledges the support 
by the project {\sl P14984} of the FwF.
Use was made of the WEBDA database, operated at the University
of Vienna, the NASA's Astrophysics Data System and
data products from the Two Micron All Sky Survey, 
which is a joint project of the University of Massachusetts and the Infrared 
Processing and Analysis Center/California Institute of Technology, funded by the 
National Aeronautics and Space Administration and the National Science Foundation. These observations have been funded by the
Optical Infrared Coordination network (OPTICON),
a major international collaboration supported by the
Research Infrastructures Programme of the
European Commission's Sixth Framework Programme.
\end{acknowledgements}

\end{document}